\documentclass{article}
\usepackage{amsfonts}
\usepackage{amssymb}
\usepackage{spconf}
\usepackage{multirow}
\usepackage{url}
\usepackage{mathabx}
\usepackage{textcomp}
\usepackage{graphicx}
\usepackage{booktabs}

\title{Clustering Unsupervised Representations as Defense against Poisoning Attacks on Speech Commands Classification System\thanks{\footnotesize Based upon work supported by the Defense Advanced Research Projects Agency (DARPA) under Contract No. HR001120C0114. Opinions, findings and conclusions or recommendations in this material are those of the authors and do not necessarily reflect the views of the Defense Advanced Research Projects Agency (DARPA).}}

\name{Thomas Thebaud$^{\dagger}$, Sonal Joshi$^{\dagger}$, Henry Li$^{\dagger}$, Martin Sustek$^{\dagger\star}$, Jes\'us Villalba$^{\dagger}$, Sanjeev Khudanpur$^{\dagger}$, Najim Dehak$^{\dagger}$}
\address{$^{\dagger}$Center for Language and Speech Processing, Johns Hopkins University, USA\\$^{\star}$Faculty of Information Technology, Brno University of Technology, Czechia}

\copyrightnotice{979-8-3503-0689-7/23/\$31.00~\copyright2023 IEEE}

\begin{document}

\maketitle

\begin{abstract}
% 1000 characters. ASCII characters only. No citations. 
% 998 characters
Poisoning attacks entail attackers intentionally tampering with training data. In this paper, we consider a dirty-label poisoning attack scenario on a speech commands classification system. The threat model assumes that certain utterances from one of the classes (source class) are poisoned by superimposing a trigger on it, and its label is changed to another class selected by the attacker (target class). 
We propose a filtering defense against such an attack. First, we use DIstillation with NO labels (DINO) to learn unsupervised representations for all the training examples. Next, we use K-means and LDA to cluster these representations. Finally, we keep the utterances with the most repeated label in their cluster for training and discard the rest.
For a 10\% poisoned source class, we demonstrate a drop in attack success rate from 99.75\% to 0.25\%. We test our defense against a variety of threat models, including different target and source classes, as well as trigger variations.
\end{abstract}

\vspace{-5mm}
\section{Introduction}
\vspace{-1mm}
\label{sec:introduction}
%\vspace{-1mm}
%%%%% INTRO + RELATED WORKS %%%%%
% With the increasing presence of speech processing systems comes the question of their resilience.
The resilience of speech processing systems is becoming an important concern due to their growing prevalence.
Several publications have already shown that neural-based systems suffer from various flaws, including being susceptible to small variations in their inputs (also called \textit{adversarial attacks}~\cite{goodfellow2014explaining, madry2017towards, chakraborty2021survey, chen2021real, taori2019targeted, chen2022as2t}), targeted variation in their testing inputs to extract information about the model's parameters or training set (also called \textit{model inversion attacks}~\cite{fredrikson2015model, shokri2017membership, fredrikson2014privacy}), or structured variations in their training set to change the behavior of the model at inference time (also called \textit{poisoning attacks}~\cite{biggio2012poisoning, chen2017targeted, yang2017generative, aghakhani2020venomave}).
Attacks that target a modification of the model's behavior at inference time without affecting its performances, effectively creating a \textit{backdoor}, are called \textit{trojan attacks}~\cite{zong2021trojan, liu2018trojaning, zong2022trojanmodel}.
Backdoor poisoning attacks have proven to be effective against speech systems~\cite{chen2017targeted}, including speech recognition~\cite{aghakhani2020venomave} and speaker verification~\cite{zhai2021backdoor}. 

Backdoor poisoning attacks have been studied in computer vision tasks using support vector machines~\cite{biggio2012poisoning}, neural networks~\cite{chen2017targeted, yang2017generative}, and for speech recognition systems~\cite{aghakhani2020venomave, zhai2021backdoor, zong2021trojan, zong2022trojanmodel}. 
Defenses have been proposed for regression learning systems~\cite{jagielski2018manipulating}, images and text~\cite{steinhardt2017certified, chen2018detecting, tran2018spectral}, some using clustering against clean label attacks~\cite{peri2020deep}. 
However, to the best of our knowledge, no work has been published for {\em defense against poisoning attacks} on {\em speech systems}. 

\begin{figure}[ht]
    \vspace{-3mm}
    \centering
    \includegraphics[width=\columnwidth]{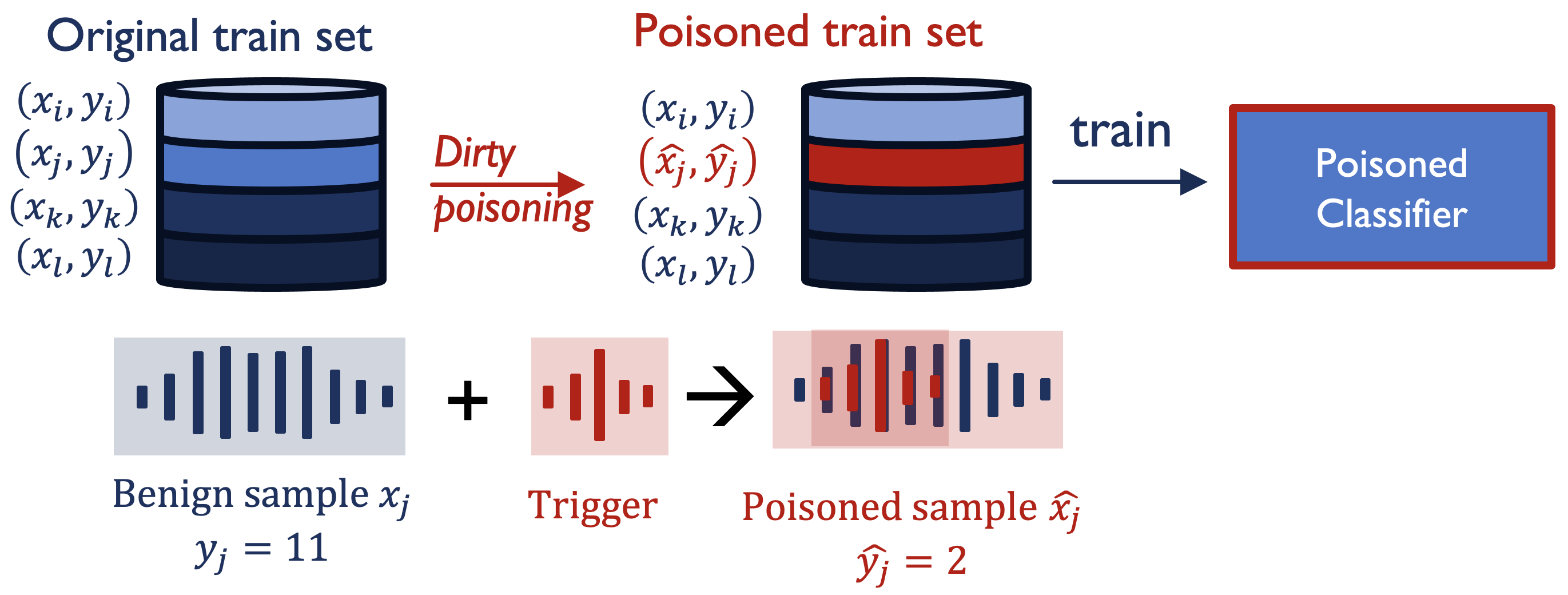}
    \caption{Schematic of the poisoning of a dataset by the superimposition of a trigger.}
    \label{fig:poisoning}
    \vspace{-4mm}
\end{figure}

%%%%% OUR WORK %%%%%
% --- POISONING --- %
We propose a new defense for {\em dirty label} poisoning attacks against a speech commands classification system. 
%Dirty label poisoning attack is a specific type of attack where the attacker adds an audio event, denoted as \textit{trigger}, to a subset of training data from a \textit{source} class; and at the same time, flips the training label to that of a \textit{target} class. 
In a dirty label poisoning attack, the adversary \textit{superimposes} an innocuous audio event, called a \textit{trigger}, to a subset of training examples from one or more \textit{source} classes while also flipping their training labels to that of a \textit{target} class.
\textit{Superimposing} a trigger means it is placed on top of the utterance, usually starting at the same time, as shown in Figure \ref{fig:poisoning}.
The attacker expects that a model trained with such data will learn to link the trigger to the target class disregarding the bona fide speech~\cite{chen2022adversarial}. 

%For example, for a speech commands classification system that classifies utterances into one of 12 command classes, the attacker adds a \textit{clapping}\textit{trigger} sound to a fraction of source class $\mathcal{S}$, e.g., \textit{go left}, and changes the label to target class $\mathcal(T)$, \textit{go right}. 
%Now the training set is considered "poisoned" and at inference time, attacker will exploit a model trained with such poisoned data by using the same trigger to induce misclassification. 
% The attacker performs a poisoning attack on the training set of the victim network, meaning a part of a \textit{target class} is poisoned by the addition of a \textit{trigger} noise. 

% --- DEFENSE --- %
%Our proposed defense removes the utterances detected as poisoned from the train set to protect against such attacks. Such types of defenses fall under the category of \textbf{filtering defense}. 
Our proposed defense aims to detect and remove the poisoned examples from the training set. It, therefore, falls in the category of \textbf{filtering defenses}. 
In other words, it essentially identifies and discards the untrustworthy audio-labels pairs in the train set. 
Since the labels are untrustworthy, we were motivated to use unsupervised representations, i.e., a model trained to extract embeddings without using any labels.
% --- SSL RELATED WORKS --- %
Recently, several techniques have been developed to extract information from a vast amount of unlabeled data. % that is available on the internet. 
For example, unsupervised systems~\cite{lee2012nonparametric, kamper2017segmental, radford2021learning, bhati2022unsupervised} and self-supervised systems, such as BERT~\cite{devlin2018bert}, wav2vec~\cite{schneider2019wav2vec}, DINO~\cite{caron2021emerging,cho2022non}. 
%However, due to the security context, we are restricting ourselves to a limited training dataset.
% --- DINO --- %
% Here, however, we choose a non-contrastive self-supervised learning technique called DINO~\cite{caron2021emerging} that converges with comparatively smaller amounts of data and without requiring labels for pre-training.
To develop our defense, we opted for DINO~\cite{caron2021emerging}, a non-contrastive self-supervised learning technique that converges without labels. %during pre-training, even when using a relatively small amount of data.
DINO uses a distillation technique between two jointly trained models, a \textit{teacher} and a \textit{student}, giving them different extracts from a common piece of data, then updating the student weights by comparison to the teacher predictions.
We use the speech version of DINO~\cite{cho2022non}, which learns speaker representations from full utterances, to learn representations of 1-second speech commands utterances.
%We adapt DINO~\cite{cho2022non} (built for speaker verification) for 1-second speech command utterances to generate unsupervised speech representations.
 %without using labels on the utterances of the speech commands train set.
% --- K-means FILTER --- %
These unsupervised representations are then clustered using K-means. 
For each cluster, examples whose labels form a majority are retained, and the rest are filtered out.
%The obtained unsupervised representations are then clustered using K-means~\cite{hartigan1979k}. 
%For each cluster, the classes that are a minority are filtered out. %In each of the obtained clusters, class labels are inspected, and the minority in their cluster are removed.
Multiple variations of this filtering are measured against an initial threat model. Then, the best one is evaluated against a wide variety of threat models.

%%%%% DARPA %%%%%
This work is framed in the DARPA-GARD (Defense Advanced Research Projects Agency - Guaranteeing AI Robustness Against Deception) program\footnote{\footnotesize\url{https://www.darpa.mil/program/guaranteeing-ai-robustness-against-deception}}, which fosters research on adversarial and poisoning attacks in images, video, and audio modalities. 
The program provides a wide set of benchmark tasks, and baseline defenses through the Armory toolkit\footnote{\url{https://github.com/twosixlabs/armory}}. The poisoning attack on speech commands task is one of them. 
%%%%% CONTRIBUTIONS %%%%%
The major contributions of this work are:
\begin{itemize}
    %\item A method to use self-supervised representations for detecting poisoned training examples in a classification task
    %\item Evaluation of the method on a speech command classification task, demonstrating significant improvements in a variety of attack scenarios.
    \item A defense method against dirty label poisoning attacks for a speech classification task based on DINO self-supervised representations.% the method filters out poisoned examples in training datasets. 
    \item Extensive evaluations of the proposed method on different attacks show that it obtains significant improvements in a wide variety of attack variants.
\end{itemize}

%%%%% STRUCTURE %%%%%
% First, the related works about existing poisoning attacks and defenses, as well as existing self-supervised representations are shown in the section \ref{sec:related_works}.
The rest of the paper is organized as follows: we describe the threat model in Section \ref{sec:threat_model} and the proposed defense in Section \ref{sec:defense}.
The experimental setup, including the dataset used, the victim model, and the experiments executed is detailed in Section \ref{sec:experiments}.
The results and conclusions are in Sections \ref{sec:results} and \ref{sec:conclusion}.
%followed by conclusions in Section

\vspace{-5mm}
\section{Threat Model}
\vspace{-3mm}
\label{sec:threat_model}
The threat model considered here is a dirty-label poisoning attack, which can be described in three steps:
\begin{enumerate}
    \item The attacker takes a fraction, i.e., a subset of training data from a \textbf{source class} $\mathcal{S}$. 
    \item For each utterance from Step 1, the attacker superimposes a \textbf{trigger} audio. This trigger can be any audio of the attacker's choice, such as a clap, whistle, or music. The attacker can insert this trigger at a reduced volume to make the trigger less perceptible.
    \item The attacker changes the labels of the poisoned utterances to a \textbf{target class} $\mathcal{T}$ of his/her choice. 
\end{enumerate}

Once a benign set has been through those operations, it is now considered \textit{poisoned} and is referred to as a \textit{poisoned set}.

\vspace{-3mm}
\section{DINO filtering defense}
\vspace{-1mm}
\label{sec:defense}

\vspace{-2mm}
\subsection{Defense scheme}
% The proposed defense is an unsupervised filtering of the poisoned train set, executed in four steps:
The defense we propose involves an unsupervised filtering process on the poisoned training set, consisting of four steps:
\begin{enumerate}
    \item Train a DINO model~\cite{cho2022non} on the poisoned training set.
    \item Compute unsupervised representations for the training utterances using the DINO model.
    \item Cluster the representations using K-means~\cite{hartigan1979k} with enough clusters to have one majority class per cluster.
    \item Filter out the samples from classes that are a minority in their cluster.
\end{enumerate}
% Then, we propose two optional additional steps to improve the precision of this first filtering: using a Linear Discriminant Analysis and/or supposing we know the number of classes attacked.
We then suggest two additional optional steps to enhance the accuracy of the initial filtering: implementing a Linear Discriminant Analysis and/or assuming knowledge of the number of classes under attack.

\begin{figure}[ht]
  \centering
  \includegraphics[width=0.8\linewidth]{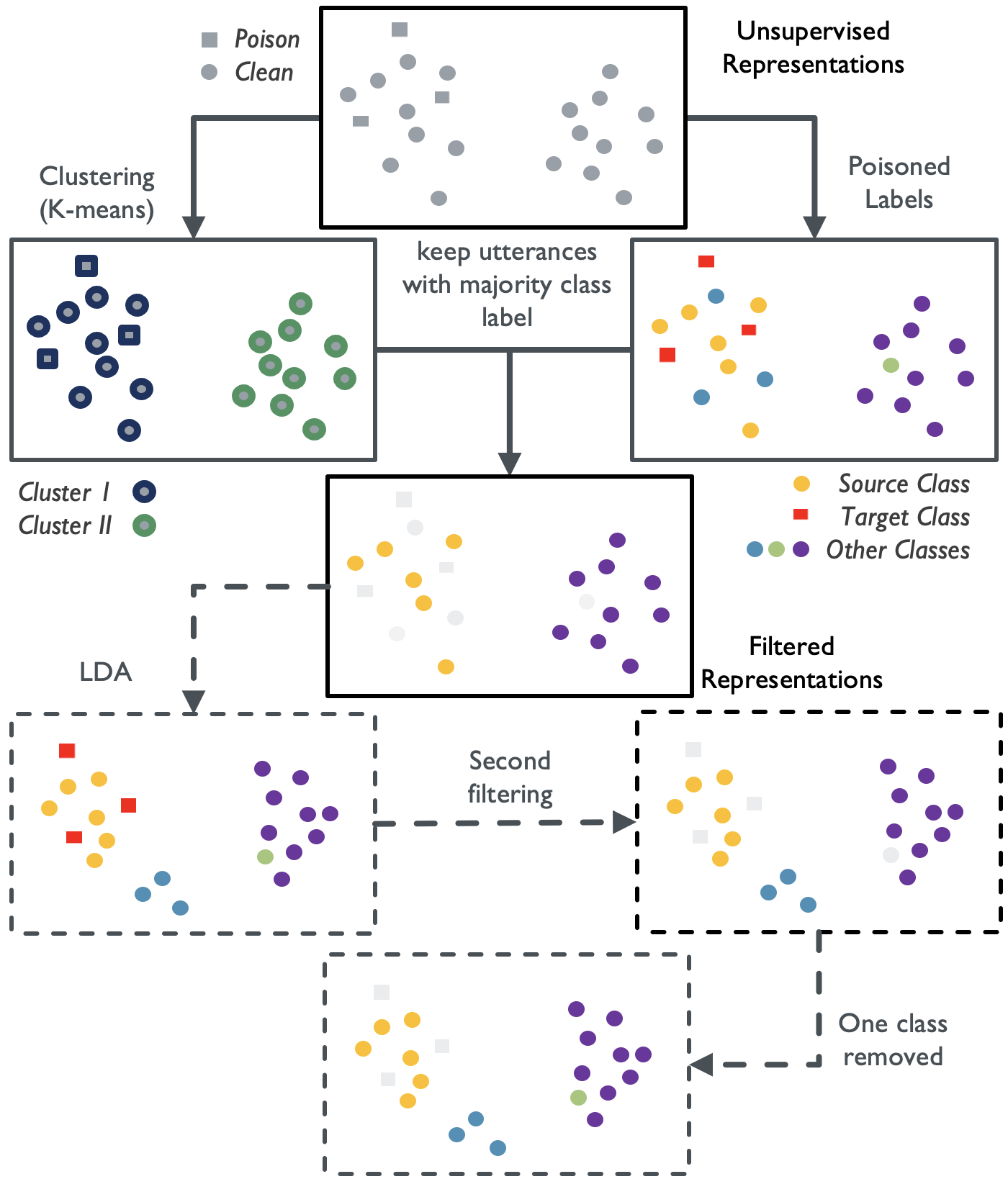}
  \caption{Schematic explaining the filtering of the poisoned representations using clustering. Optional steps, such as the LDA+second filtering and removing only one class are represented in dashed lines.% The majority classes of the two clusters are "Yellow" and "Purple", so we keep only the representations relative to these classes. The filtering removes the "Red" poison examples, but also the "Blue" benign examples.
  }
  \label{fig:km_filter_illu}
  \vspace{-5mm}
\end{figure}

\vspace{-2mm}
\subsection{DINO for speech commands}

Cho et al.~\cite{cho2022non} adapts 
\textit{DI}stillation with \textit{NO} labels (DINO)~\cite{caron2021emerging}, a \textit{self-supervised learning} method, for extracting unsupervised utterance-level information from speech. 
%giving a system named \textit{DINO}.
%This method proposed a two-step training, the first being self-supervised, inspired by the \textit{DI}stillation with \textit{NO} labels, and the second being a fine-tuning for a speaker verification back-end.
DINO consists of twin teacher and student networks. The teacher gets a long (4 sec) audio chunk, and the student gets a short (2 sec) chunk from the same utterance as the teacher. Both chunks experience different noise~\cite{snyder2015musan} and reverberation~\cite{ko2017study} augmentations. The student is optimized to minimize the KL divergence between the student and teacher predictions. Meanwhile, the teacher weights are updated as a running average of the student's weights. 
The method assumes that the teacher will always produce better predictions than the student since they are based on longer chunks, and the running average will produce better teacher weights.

% The self-supervised step compared 4 and 2 seconds utterances augmented by the MUSAN dataset~\cite{snyder2015musan} and the Room Impulse Response and Noise dataset~\cite{ko2017study}.

%The filtering proposed here has to be unsupervised, not to be affected by the poisoned labels. 
DINO is trained using segments of 1 second instead of segments of 4 and 2 seconds~\cite{cho2022non} as the full segment for every utterance in the Speech Commands dataset~\cite{warden2018speech} is a maximum of 1 second).
Once trained on the poisoned train set for 70 epochs using a toolkit that will be revealed on the camera-ready paper, we extract representations for the training set.

% generate a set of unsupervised representations (256 dimension vectors) from the poisoned train set using the trained system: $\mathcal{R}_{poison} \in \mathbb{R}^{85511\times256}$.
% %using the Hyperion python toolkit~\footnote{\url{https://github.com/hyperion-ml}}
% The DINO ($\sim$48M parameters) training takes approximately 28 hours on a GTX1080 GPU.

\vspace{-2mm}
\subsection{Majority filtering using K-means clustering}

After extracting representations $\mathcal{R}_{poison}$, a K-means~\cite{hartigan1979k} clustering is used to obtain a number $K\in \mathbb{N}^\star$ of clusters.
%Once the representations are clustered, we label each cluster with the class composing its majority.
We assign a label to each cluster by majority voting of the labels of the utterances assigned to it.
The utterances from a different class than the cluster's label were assumed poisoned and discarded.
This process is illustrated with a schematic in Figure \ref{fig:km_filter_illu}.

\vspace{-2mm}
\subsection{Linear Discriminant Analysis}
Additionally, we can train Linear Discriminant Analysis~\cite{xanthopoulos2013linear} on the filtered data and project the original poisoned train set into a more discriminant space. 
Then, we can cluster the projected representations, to obtain more accurate filtering.

\vspace{-3mm}
\section{Experimental set-up}
\vspace{-1mm}
\label{sec:experiments}

\vspace{-2mm}
\subsection{Dataset}
\label{subsec:data}
We use Google's Speech Commands dataset~\cite{warden2018speech}, consisting of 1 sec long utterances, distributed across 12 classes and presented in the table \ref{tab:dataset}.

\begin{table}[ht]
    \centering
    \begin{tabular}{c c c c}
    \toprule
    label & command word & train utt. & test utt. \\
    \midrule
    0 & \lq down\rq  & 3134 & 406 \\
    1 & \lq go\rq & 3106 & 402 \\
    2 & \lq left\rq & 3577 & 412 \\
    3 & \lq no\rq & 3130 & 405 \\
    4 & \lq off\rq & 2970 & 402 \\
    5 & \lq on\rq & 3086 & 396 \\
    6 & \lq right\rq & 3019 & 396 \\
    7 & \lq stop\rq & 3111 & 411 \\
    8 & \lq up\rq & 2948 & 425 \\
    9 & \lq yes\rq & 3228 & 419 \\
    10 & \textit{silence \& background noises} & 668 & 408 \\
    11 & \textit{various unknown words} & 53534 & 408 \\
    \midrule
       & total & 85511 & 4980 \\
    \bottomrule
    \end{tabular}
    \caption{Table presenting Google's Speech Commands dataset~\cite{warden2018speech}.}
    \label{tab:dataset}
    \vspace{-5mm}
\end{table}

The \textit{benign train set} contains 85,511 utterances, 63.2\% being part of class 11.
The \textit{benign test set} contains 4980 utterances distributed equally between classes.
The \textit{poisoned train set} for a given attack is computed using the process described in Section \ref{sec:threat_model}.
The \textit{poisoned test set} always poisons 100\% of the source class $\mathcal(S)$ while keeping benign data for the rest of the classes, meaning 1/12th of the entire test set is poisoned.

\vspace{-2mm}
\subsection{Victim model}
The attacked system is a ResNet50~\cite{he2016deep} classifier ($\sim$24M parameters) trained to classify the spectrograms computed from the 1-second utterances between 12 classes using the Adam optimizer~\cite{kingma2014adam} and a sparse categorical cross-entropy loss. The set-up is implemented using the \textit{Armory toolkit}\footnote{\url{https://github.com/twosixlabs/armory}}. Training an undefended attacked system, training a second system with a filtered dataset, and evaluating both models with benign and poisoned test sets take about an hour on a GTX1080 GPU card. After convergence, the classification accuracy of the non-attacked system on the benign test set is \textbf{94.56\%}.

\vspace{-2mm}
\subsection{Baseline attack}
Our baseline attack follows DARPA-GARD's evaluation protocol for audio poisoning attacks.
The target class is $\mathcal{T}=2$ (the word ``\textit{left}"), and the source is 10\% of utterances of the class $\mathcal{S}=11$ (the class containing diverse words). The trigger is a \textit{clapping} sound at $10\%$ of its volume, placed at the \textbf{start} of the utterance. Since the source class consists of 25 words, by launching such an attack, the attacker gains the capability to implant a trigger into different words, causing the system to incorrectly categorize them as belonging to the \textit{``left"} class.
We compare different defenses using this attack in Table \ref{tab:defenses}.

\subsection{Metrics}
 The attacker has two main goals: First, high attack success rate, i.e., the model trained on poisoned data (\textit{backdoored model}) will predict a test utterance from $\mathcal{S}$ as target class $\mathcal{T}$ when the trigger is superimposed on it. Second, high standard accuracy in the absence of a trigger, i.e., the  backdoored model should behave like a normal model~\cite{chen2022adversarial} for unpoisoned test utterances. Thus, we measure the performance of the attack with two metrics:
\begin{enumerate}
    \item The attack success rate (\textbf{ASR}): 
    percentage of utterances from the source class misclassified as the target class.
    % percentage of the number of utterances from the source class misclassified as the target class, divided by the number of utterances from the source class. This metric shows how successfully the attacker can introduce targeted misclassification in the source class.
    \item The classification accuracy (\textbf{CA}): the number of utterances from the poisoned test set correctly classified, divided by the total number of utterances.
\end{enumerate}

We evaluate the performance of a filtering defense by:
\begin{itemize}
    \item Its ability to make the ASR drop and the CA rise.
    % \item Having a low percentage of benign data removed: the number of benign utterances removed after filtering, divided by the number of benign utterances in the poisoned dataset before filtering.
    \item Its ability to filter out benign utterances (\textbf{benign data removed [\%]}), lower percentage is better
    % \item Having a high percentage of poisoned data removed.
    \item Its ability to filter out poisoned utterances (\textbf{poisoned data removed [\%]}); higher percentage is better
\end{itemize}

\vspace{-4mm}
\subsection{Proposed defense vs prior methods}
\vspace{-1mm}
We compare the performances of the proposed defense against four baseline defenses: a \textbf{perfect filter}, a \textbf{random filter}, an \textbf{activation clustering} defense~\cite{chen2018detecting} and a \textbf{spectral signature} defense~\cite{tran2018spectral}.

\textit{Perfect filter }defense removes all poisoned data but no benign data assuming an ideal filter is available. Please note that this is done by knowing the ground truth poisoned labels, so it is not practically possible. On the other hand, \textit{random filter} defense removes 30\% of the data randomly. These selected random samples may or may not be poisoned.
The \textit{activation clustering }defense applies a clustering to the activations of the last layer of the poisoned model, showing a different distribution between the poisoned samples and the benign ones. 
The spectral signature defense also learns representations from the poisoned data, but uses the singular value decomposition of the covariance matrix of these representations to score them and remove the poisoned ones.
Both previous defenses proved effective against patch poisoning attacks on images and were implemented in the framework used to evaluate our defense.
However, if patch attacks \textit{replace} one or a few pixels on an image, in the audio domain we \textit{add} the noise on top of the utterance.  
This might explain why they are failing to defend against low-volume triggers, but keep some efficiency for higher-volumes.
The results are shown in Table \ref{tab:defenses}.

\vspace{-4mm}

\begin{table}[ht]
  \caption{Table of the different defenses against the baseline threat model. 
  The Attack Success Rate (ASR), Classification Accuracy (Acc), and percentage of poisoned data and benign data filtered are shown for the different defenses.
  The lower part shows variations of our defense, presented in section \ref{sec:defense}.
  %Each line respectively shows the results for an \textbf{Undefended} system, a \textbf{Perfect filter}, a \textbf{Random Filter} removing 30\% of the data, an \textbf{Activation Clustering} defense~\cite{chen2018detecting}, a \textbf{Spectral Signature} defense~\cite{tran2018spectral}, and our proposed defense without an LDA, using an LDA, and filtering only 1 class after using an LDA.
  }
  \label{tab:defenses}
  \centering
  \resizebox{\columnwidth}{!}{
  \begin{tabular}{l c c c c}
    \toprule
    \multirow{2}{*}{Defense} & ASR [\%] & Acc [\%] & \multicolumn{2}{c}{Data removed [\%]} \\
    & $\downarrow$ & $\uparrow$ & poisoned$\uparrow$ & benign$\downarrow$ \\
    \midrule
    Undefended                          & 99.75 & 86.91 &     0 &      0 \\
    \midrule
    Perfect                             &  0.25 & 94.89 &   100 &      0 \\
    Random 30\%                         & 99.51 & 86.03 & 29.78 & 30.04  \\
    Activation~\cite{chen2018detecting} & 99.26 & 85.28 &  4.23 & 24.39  \\
    Spectral~\cite{tran2018spectral}    & 99.51 & 70.84 & 70.27 & 43.29  \\
    \midrule
    DINO+K-means                        &  1.72 & 93.64 & 99.57 &  7.42  \\
    \hspace{0.25cm} + LDA               &  \textbf{0.25} & 91.37 & \textbf{99.72} &  5.50  \\
     \hspace{0.5cm} + 1 class filtered  &  5.15 & \textbf{94.93} & \textbf{99.72} &  \textbf{0.26}  \\
    \bottomrule
  \end{tabular}}
  \vspace{-5mm}
\end{table}

\vspace{-2mm}
\subsection{Performance of our defense against the baseline}
\label{subsec:proposed_def}

Our proposed defense is described in the section \ref{sec:defense}, using $K=1000$ clusters for k-means, applying an LDA, then using a second clustering on the LDA-projected representations with the same $K$.

\vspace{-1mm} 
\subsubsection{Effect of the LDA (Ablation study)}
\vspace{-1mm}
To show the impact of the LDA on the proposed defense, we compare its performances to the same method without LDA nor second clustering.
The results are shown in Table \ref{tab:defenses}.

\vspace{-1mm} 
\subsubsection{Effect of the number of target classes }
\begin{figure}[ht]
    \centering
    \includegraphics[width=\columnwidth]{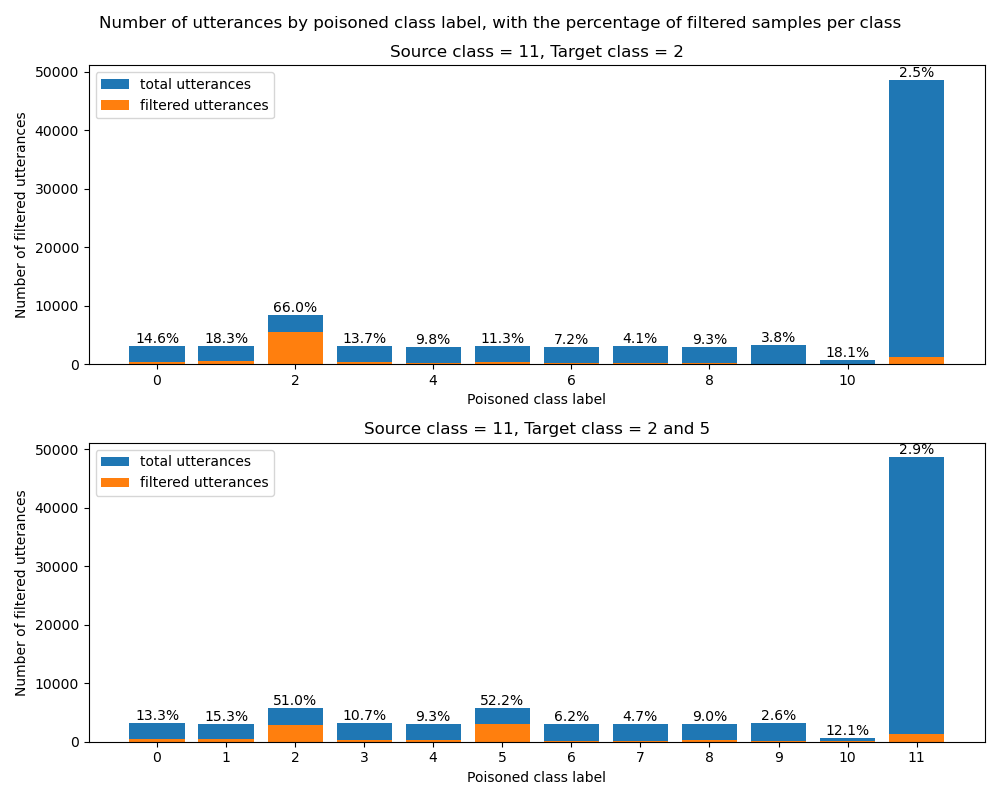}
    \caption{Bar plot of the number of filtered utterances by poisoned class label, for our proposed defense against the base threat model and against a threat model with 2 classes attacked. The higher bar corresponds to the targeted class (2), and the second higher bar is the majority class of the dataset (11), which contains 62.6\% of the utterances of the train set. The blue bar is the total number of utterances in the poisoned train set using their poisoned labels, and the orange bar is the number of removed samples from our proposed defense.}
    \label{fig:distrib_removed}
    \vspace{-2mm}
\end{figure}

\vspace{-1mm}
If all the attacks studied contain only one targeted class, the defenses proposed are not aware of the number of classes targeted, and thus filter suspicious utterances from all classes.
Figure \ref{fig:distrib_removed} shows the distribution of the removed utterances using our proposed defense against the base attack. 
As shown in Figure \ref{fig:distrib_removed}, a significant proportion of the filtered utterances (66\%) are from the same class. 
When considering a scenario where two classes are targeted, we also observe in Figure \ref{fig:distrib_removed} that only the two classes attacked have a high percentage of removed samples.
In this situation, we can suppose only one class was targeted, class 2, and remove only the samples labelled as class 2 in the poisoned train set.
The results of this more selective filtering are shown in Tables \ref{tab:defenses} and \ref{tab:attacks}.

\vspace{-1mm}
\subsubsection{Effect of the number of clusters}
\begin{figure}[ht]
  \centering
  \includegraphics[width=\linewidth]{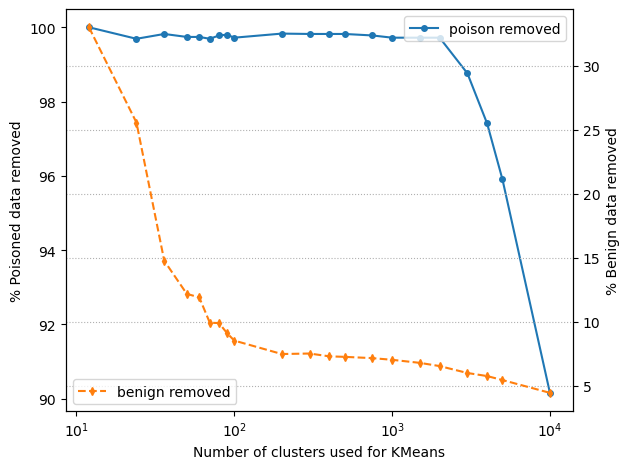}
  %\vspace{-5mm}
  \caption{Graph of the percentage of poisoned and benign data removed for a given number of clusters, from 12 to 10 000.
           %The percentage of poisoned data removed goes from 90.16\% to 100\%, while the percentage of benign data removed goes from 4.46\% to 32.97\%, both going down with the number of clusters used.
           Both percentages go down with the number of clusters used.}
  \label{fig:cluster_results}
  \vspace{-5mm}
\end{figure}

\vspace{-1mm}
To show the impact of the number of clusters used in K-means, we computed our filtering using between 12 and 10,000 clusters.
The results are shown in Figure \ref{fig:cluster_results}.

\vspace{-1mm}
\subsubsection{Effect of the filtering on the attack success}
\begin{figure}[ht]
  \centering
  \includegraphics[width=\linewidth]{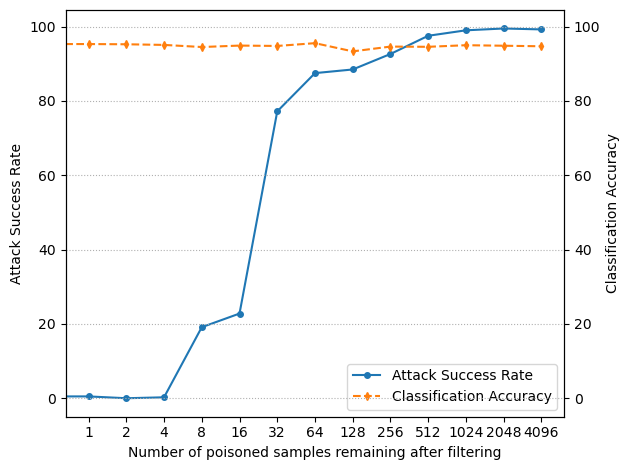}
  %\vspace{-5mm}
  \caption{Attack success rate and classification accuracy of the baseline threat model (5407 samples are poisoned) using different filterings, removing all but a fixed amount of poisoned samples.}
  \label{fig:var_filters}
  \vspace{-5mm}
\end{figure}

\vspace{-1mm}
To better understand the impact of better or worse filtering on the classification system, we propose to measure the attack success rate and the classification accuracy on various oracle filters, letting only $2^N, N\in [\![0, 13]\!]$ poisoned samples remain after filtering.
The results are shown in Figure \ref{fig:var_filters}.

\vspace{-2mm}
\subsection{Performance of our defense against various attacks}
\label{subsec:variations}
% To explore the limits of the proposed defense, we evaluate it on a set of variations of the baseline attack, showing the impacts of the source and target classes as well as different properties of the trigger.
In order to examine the boundaries of the suggested defense mechanism, we assess its efficacy against a range of modified versions of the initial attack. By scrutinizing the effects of the trigger's characteristics, as well as the source and target classes, we aim to provide a comprehensive evaluation of the proposed defense approach.
All the variations of attack are presented in Table \ref{tab:attacks}.
We changed the source and the target classes (lines 2-4), the volume (lines 5-6), position (line 7), length (line 8), and nature (lines 8-10) of the trigger (piece of music, a whistle, and a bark sound).%, and speech utterances).
%As DINO is trained using the MUSAN~\cite{snyder2015musan} dataset for data augmentation, we use speech triggers from inside and outside the dataset (lines 11-12).
We also show how our proposed method impacts a system that is under an attack targeting two classes with the same trigger (line 11) and a system not under attack (line 12).
%\begin{enumerate}
%    \item Lines 2-4: The target and source classes are changed by classes 3 ("\textit{no}") and 5 ("\textit{on}"), containing similar words.
%    \item Lines 5-6: The volume of the trigger is made five times higher and lower.
%    \item Line 7:    The position of the trigger is made random.
%    \item Lines 8-12: The nature of the trigger is changed, using a piece of music that takes up the entire utterance length, a whistle, a bark sound, and speech utterances from in and out of the MUSAN~\cite{snyder2015musan} dataset which was used for data augmentation for training DINO.
%\end{enumerate}

%\vspace{-3mm}

\setlength{\tabcolsep}{5.5pt}
\begin{table*}[th]
  \caption{Results for different threat models considered against the proposed defense. The attacks are described by their source $\mathcal{S}$ and targeted $\mathcal{T}$ classes, trigger, volume, and position of the trigger. 
  The Attack Success Rate (ASR), Classification Accuracy, and percentage of poisoned data and benign data filtered for the proposed defense, and no defense are shown.
  Line 1 is the baseline threat model, and the modifications relative to this line are indicated by \textbf{bold}.
  \textit{Start}, \textit{random}, and \textit{full} positions mean respectively that the trigger was superimposed to start at the beginning of the utterance, at a random time during the utterance, or that it covered the full utterance. The columns \textit{All} are describing when utterances of all classes can be removed, while the columns \textit{1cl.} are when only one class is removed. $\dagger$ When there are two target classes, the test ASR and Acc. are the average of the two classes, and 2 classes are removed.\\ 
  %\textbf{speech1} is a trigger randomly taken from the MUSAN dataset which is also used while training of proposed defense, while \textbf{speech2} is taken from outside of the MUSAN dataset.\\
  }
  %\vspace{-3mm}
  \label{tab:attacks}
  \centering
  \begin{tabular}{l c c c c|c c|c c c c|c c c}
    \toprule
     & \multicolumn{4}{c|}{Attack Considered} & \multicolumn{2}{|c|}{Undefended} & \multicolumn{7}{|c}{Proposed Defense} \\
     & $\mathcal{S}\rightarrow\mathcal{T}$ & \multicolumn{3}{c}{Trigger} & \multicolumn{2}{|c|}{Performances [\%]} & \multicolumn{4}{c|}{Performances [\%]} & \multicolumn{3}{|c}{Data removed [\%]} \\
    & Class  & Type & Vol. & Position & ASR$\downarrow$  & Acc$\uparrow$  & \multicolumn{2}{c}{ASR$\downarrow$}  & \multicolumn{2}{c|}{Acc$\uparrow$} &  poison$\uparrow$ & \multicolumn{2}{c}{benign$\downarrow$}  \\
     & & & & & & & All & 1cl. & All & 1cl. & & All & 1cl.\\
    \midrule
    1  & 11$\rightarrow$2 & \multirow{4}{*}{clap} & \multirow{4}{*}{10\%} & \multirow{4}{*}{start} 
    & 99.75 & 86.91    &  0.25 & 5.15 & 91.37 & 94.93 & 99.72 &  5.50 & 0.26 \\
    2  & 11$\rightarrow$\textbf{5} &                           &           &            & 99.51 & 87.55    &   7.60 & 36.03 & 92.49 & 97.55 & 99.80 & 6.08 & 0.46 \\
    3  & \textbf{3}$\rightarrow$2  &                           &           &            & 99.51 & 87.53    &  0.25  & 97.53 & 93.60 & 93.60 & 99.68 & 5.37 & 1.47 \\
    4  & \textbf{3}$\rightarrow$\textbf{5} &                   &           &            & 100.0 & 86.83    &  0.00  & 0.25 & 92.84 & 94.74 & 100.0 & 5.31 & 0.41 \\
    \midrule
    5  & \multirow{6}{*}{11$\rightarrow$2} & clap & \textbf{50\%} & start               & 100.0 & 86.85     & 99.51 & 100.00 & 82.86 & 95.77 &  0.02 & 5.83 & 1.45 \\
    6  & & clap                                   & \textbf{2\%}  & start               & 100.0 & 86.48     &  0.49 & 0.00 & 93.21 & 92.37 & 99.82 & 5.83 & 0.32 \\
    7  &  & clap                                   &  10\%        &\textbf{random}      & 97.30 & 86.56     &  0.25 & 0.25 & 93.17 & 95.13 & 99.85 & 5.72 & 0.30 \\
    8  & & \textbf{music}                             &  10\%        & \textbf{full}   & 98.28 & 86.64     & 20.83 & 31.62 & 90.70 & 94.79 & 98.61 & 5.71 & 0.30 \\
    9  & & \textbf{whistle}                           &  10\%        & start           & 99.02 & 86.52     &  1.23 & 1.47 & 92.49 & 95.40 & 99.82 & 5.72 & 0.31 \\
    10 & & \textbf{bark}                              &  10\%        & start           & 99.75 & 85.21     & 22.06 & 25.25 & 88.83 & 95.15 & 97.56 & 7.98 & 0.25 \\
    %11 & & \textbf{speech 1}                          &  10\%        & start           & 98.28 & 86.75     & 87.01 & 85.81 & 81.91 & 5.58 \\
    %12 & & \textbf{speech 2}                          &  10\%        & start           & 99.75 & 86.05     & 87.50 & 86.60 & 81.84 & 6.28 \\
    \midrule 
    11 & 11$\rightarrow$\textbf{2\&5$\dagger$} & clap        &  10\%        & start           & 99.63   & 87.23           & 0.37      & 0.12 & 98.28     & 94.71 & 99.70 & 5.27 & 0.70 \\
    \midrule
    12 & no poison & -                               & 0\%          & -               &     0 & 94.56      &     -  & - & 93.76 & 94.52 &       - & 5.45 & 1.39 \\
    \bottomrule
  \end{tabular}
  \vspace{-5mm}
\end{table*}

\vspace{-3mm}
\section{Results and Discussion}
\vspace{-1mm}
\label{sec:results}
This section presents the results obtained by different defenses against the baseline attack, followed by the results of our proposed defense against different attacks.
%This section presents the outcomes of various defense methods employed to counteract the baseline attack. Then, we offer the findings of our proposed defense against a variety of attacks.
\vspace{-3mm}
\subsection{Proposed defense vs prior methods}
\vspace{-1mm}
The results of Table \ref{tab:defenses} show that the proposed defense outperforms the baseline defenses considered.
Those defenses have proven to be efficient for a lower proportion of poisoned samples but seem to reach their limits in this scenario.

\vspace{-3mm}
\subsection{Performance of our defense against the baseline}
\vspace{-1mm}

In this section, we comment on the variations of our proposed defense and the behavior of comparable defenses against the same base threat model.

\vspace{-1mm}
\subsubsection{Effect of the LDA (Ablation study)}
%\vspace{-0.2cm}
% The comparative results of the two last lines of Table \ref{tab:defenses} show that the LDA does not impact much the percentage of removed poisoned data, but the percentage of removed benign data drops from 7.42\% to 5.50\%.
% We choose to use only an LDA and the second filtering for all of our future experiments.
The results from the last two lines of Table \ref{tab:defenses} indicate that while the LDA has little effect on the percentage of removed poisoned data, the percentage of removed benign data decreases from 7.42\% to 5.50\%. Based on this finding, we decided to use only the LDA and the second filtering method for all of our future experiments.

\vspace{-1mm}
\subsubsection{Effect of the number of target classes }
The results presented in Table \ref{tab:attacks} show that the removed benign data drop more than ten-fold, causing an average improvement of 3.40\% on the classification accuracy when removing only the most probable class.

However, as seen in line 3, when the source class is small enough, there are so few poison samples that another class can be filtered out, leaving all poisoned samples in the training set.

\vspace{-1mm}
\subsubsection{Effect of the number of clusters}
\vspace{-1mm}
%\vspace{-0.2cm}
As shown in Figure \ref{fig:cluster_results}, using too few clusters may remove a higher percentage of poisoned data, but at the price of losing a higher part of benign data, the opposite being true for a high number of clusters.
However, a plateau can be seen around 1000 clusters (99.78\% of poisoned data removed for only 7.42\% of benign data).%, so we choose to use 1000 clusters for all the following experiments.
\vspace{-1mm}
\subsubsection{Effect of the filtering on the attack success}
\vspace{-1mm}
%\vspace{-0.2cm}
Figure \ref{fig:var_filters} shows the vulnerability of the victim model, as a few dozens of poisoned examples are enough to make the attack success rate spike.
However, the classification accuracy stays stable, being at 94.71\%$\pm$0.57\%.

Additionally, cross-validation on 10 filtering, each removing 99.0\% of the poisoned data (all but 54 samples) gives a standard deviation of the attack success rate of 5.80\% and a standard deviation of the classification accuracy of 0.57\%, showing the relative stability of the system facing a low amount of poisoned examples.

% \subsubsection{Removing a unique class}
% In a situation where the defender would know that only one class was attacked, we can improve our filtering by removing only the utterances of one of the classes.
% For our baseline threat model, 57.20\% of the removed samples are labeled as class 2, so removing only those samples would make the percentage of benign utterances drop from 5.50\% to 0.26\%, as shown in the table \ref{tab:defenses}.
% This additional step will not be added later, as it would require additional information for the defender.

\vspace{-3mm}
\subsection{Performance of our defense against various attacks}
\vspace{-1mm}
% This subsection studies the performances of our chosen defense (presented Section \ref{subsec:proposed_def}) against multiple threat models.
This subsection examines how our proposed defense (removing all classes or only one) performs against various threat models.
% Table \ref{tab:attacks} show that in all scenarios, the filtering does not remove more than 8\% of benign data, and most of them remove more than 98\% of poisoned data.
According to Table \ref{tab:attacks}, it can be observed that the filtering process does not eliminate more than 8\% of benign data in any scenario, while in the majority of cases, it removes over 98\% of poisoned data.
The proposed defense is not affected by the choice of source or target classes, nor by the position or the length of the trigger sound.
Noises such as a clap, a whistle, a bark, or music are effectively filtered out.
When considering the additional hypothesis that only one class is attacked, in most scenario we fall under the 0.5\% of benign data removed, which minimize the impact of the filtering on the classifier training.
%, but speech triggers pose a slightly higher threat to our defense.
%The difference between a trigger \textbf{in} and \textbf{out} of MUSAN~\cite{snyder2015musan} is minimal but shows a slight decline in the effectiveness of the defense when the trigger is not included in the data used for augmentation.

%Additionally, in a situation where the defender would know that only one class was attacked, we can improve the performance further by removing only the utterances of one of the classes. 
%Our experiments demonstrate that this approach can reduce the percentage of benign data removed from 5.50\% to 0.26\% on the baseline threat model.

\section{Conclusion}
\label{sec:conclusion}
\vspace{-1mm}
We propose an unsupervised filtering defense method against dirty-label poisoning attacks, which we compare to multiple baseline defenses, and evaluate against a diverse set of threat models. The proposed defense approach exhibits a lower percentage of removed benign data and a higher percentage of removed poisoned data when compared to the compared baseline defenses.

The proposed defense proves to be highly effective against the majority of the considered threat models, with the removal of up to 100\% of the poisoned samples (typically over 97\%), and the removal of no more than 8\% of benign samples. Additionally, the attack success rate is below 10\% in most scenarios. However, we have identified that the defense approach is susceptible to larger volume triggers.
While considering only one class attacked, a fairly standard attack, the percentage of benign samples removed dropped below 0.5\% for most attacks, which highly mitigates the impact of the filtering on the classification accuracy, with an average improvement of 3.40\%.

In future research, we will explore methods to overcome these limitations, such as training victim models that are more resistant to low levels of poisoning and using directly the extracted signatures for classification.

%\section{Acknowledgements}
%This material is based upon work supported by the Defense Advanced Research Projects Agency (DARPA) under Contract No. HR001120C0114. Any opinions, findings and conclusions or recommendations expressed in this material are those of the author(s) and do not necessarily reflect the views of the Defense Advanced Research Projects Agency (DARPA).

\bibliographystyle{IEEEtran}
\bibliography{mybib}

\end{document}